\documentstyle[aps,preprint,epsfig]{revtex}
\tightenlines

\begin{document}
\draft

\title{Population of Isomers in Decay of the Giant Dipole Resonance}
\author{N.~Tsoneva, Ch.~Stoyanov,}
\address{Institute for Nuclear Research 
and Nuclear Energy, 1784 Sofia, Bulgaria }
\author{Yu. P.~Gangrsky,
V. Yu.~Ponomarev\cite{byline1},} 
\address{Joint Institute for Nuclear Research, 
141980, Moscow region, Dubna, Russia}
\author{N. P.~Balabanov,} 
\address{P. Khilendarsky University, Plovdiv, Bulgaria}
\author{A. P.~Tonchev}
\address{Department of Physics, Idaho State University,
Pocatello, Idaho 83209, USA }
\date{\today }
\maketitle

\begin{abstract}
The value of an isomeric ratio (IR) in N=81 isotones 
($^{137}$Ba, $^{139}$Ce, $^{141}$Nd and $^{143}$Sm) is studied by means of 
the ($\gamma ,n)$ reaction.
This quantity measures a probability to populate the isomeric state in
respect to the ground state population.
In ($\gamma ,n)$ reactions, the giant dipole resonance (GDR) is excited 
and after 
its decay by a neutron emission, the nucleus has an excitation energy of a 
few MeV.
The forthcoming $\gamma$ decay by direct or cascade transitions deexcites
the nucleus into an isomeric or ground state. It has been observed 
experimentally that the IR for $^{137}$Ba and $^{139}$Ce equals
about 0.13 while in two heavier isotones it is even less than half the size. 
To explain this effect, the structure of the excited states in the energy 
region up to 6.5 MeV has been calculated within the Quasiparticle Phonon Model. 
Many states are found connected to the ground and isomeric states by $E1$, 
$E2$ and $M1$ transitions. The single-particle component 
of the wave function is responsible for the large values of the
transitions. The calculated value of the isomeric ratio is in very good
agreement with the experimental data for all isotones.
A slightly different value of maximum energy with which the
nuclei rest after neutron decay of the GDR is responsible for the reported 
effect of the A-dependence of the IR. 
\end{abstract}


%\begin{multicols}{2}
\narrowtext

\section{Introduction.}

Considerable interest in the structure of the states at intermediate 
excitation energy in atomic nuclei has arisen recently \cite
{Gang,Belov,Mazur1,Mazur2,Duben,Ponom1,Cosel,Carroll,Huber,Tonch}. 
The states are excited via different nuclear reactions and the deexcitation
populates the low-lying excited states that have a simple structure 
\cite{Gang}. 
The study of the
electromagnetic transitions coupling low-lying states with those having
intermediate energy reveals a delicate interplay between the main
excitation modes in atomic nuclei - single-particle and the collective ones 
\cite{Ponom1}.

Isomers have been known for more than 50 years and during this period
very precise spectroscopic information about their properties
has been obtained (see, e.g. a
review article \cite{Gang}).
They have relatively low excitation energy
and total angular momentum, $J_{iso}$, which is very different from the angular 
momentum of the ground state, $J_{g.s.}$.
Due to these specific properties, their electromagnetic decay 
into the ground state is strongly hindered and they are characterized 
by a half-life from ms to years depending on the value of 
$|J_{iso}-J_{g.s.}|$.  
The nuclear isomer population is studied by means of $(\gamma,\gamma')$,
$(n,\gamma)$, ($\gamma ,n)$, $(n,2n)$ reactions and $\beta $-decay. 
The different nuclear reactions show the contribution of the 
single-particle and the more complex components in the structure of their 
wave function. 

Recently, essential progress in isomer photoexcitation has been achieved
by using bremsstrahlung radiation at low values of an end-point energy
\cite{Ponom1,Cosel,Carroll,Huber}.
The isomer yield in spherical nuclei as a function of the end-point energy  
was found to have a linear dependence in the energy interval of about one MeV, 
the line then breaks, and another linear dependence for one MeV takes place.
This effect received its theoretical interpretation as follows: A level 
or a group of closely lying levels with very specific properties are located
at the line break.
The wave function of these excited states includes configurations which allow 
their intensive excitation from the ground state while some other 
configurations are responsible for a cascade decay from these levels at
intermediate energies which lead to the isomeric state
\cite{Ponom1,Cosel,Carroll,Huber}.
Usually the maximum value of the end-point energy in these bremsstrahlung 
experiments is 4--4.5 MeV.

An alternative way to populate the isomeric states is offered by the 
$(\gamma ,n)$ reaction, applying the end-point energy of the
bremsstrahlung radiation from 10 to 25 MeV.
In this reaction the giant dipole resonance (GDR) is initially excited. 
The main mechanism of its decay in the heavy nuclei is related to the 
neutron emission.
The nucleus rests with the excitation energy of 5--7 MeV and decays by
$\gamma$ transitions into the isomeric and ground states.
Since the excitation energy of intermediate states reached in this 
reaction is relatively high, the statistical approach is usually used 
for a theoretical interpretation of the $\gamma$ decay process 
\cite{Arifov,Vanden,Ignat}.

In the present paper we present the experimental results for the population
of $h_{11/2}$ isomers in N=81 isotones in the $(\gamma ,n)$ reaction. 
It will be reported that at E$_{\gamma}$ = 25 MeV the isomeric ratio (IR)
equals 0.12, 0.14, 0.06 and 0.046 in $^{137}$Ba, $^{139}$Ce, $^{141}$Nd 
and $^{143}$Sm, respectively. All these nuclei have identical spins and 
parities in their isomeric and ground states. Therefore, 
the IR should not be affected by such factors as the difference between 
the spin of the isomeric ($J^{\pi}=11/2^{-}$) and ground states 
($J^{\pi}=3/2^{+}$).
Moreover, the isomeric levels for the isotopes $^{139}$Ce, 
$^{141}$Nd and $^{143}$Sm have  an energy on the order of 755 keV and
661 keV in $^{137}$Ba. It is clear that it is not possible to explain the 
difference between two lighter and two heavier nuclei by statistical 
properties of intermediate states without additional assumptions. 
It means that the structure of the excited states is important even 
at these excitation energies, and it should be accounted for.
For this reason, a microscopic calculation should be applied to explain
the A-dependence of the isomer yield which is experimentally observed.
For that purpose, we employ the Quasiparticle Phonon Model (QPM) which has
already been successfully applied to the interpretation of the isomer 
population from the intermediate states at lower energies 
\cite{Ponom1,Cosel,Carroll,Huber}.

The structure of the states at intermediate excitation energy is 
rather complex. The interplay of the simple collective and the more complex
modes leads to the fragmentation of the simple excitations. 
The experimental data as well as theoretical calculations reveal 
the spreading of the simple modes in a wide energy region, 
where large fluctuations in strength are obtained. A
convenient approach to describe the fragmentation 
is based on the doorway picture\cite{Gale,Vdov1}. This approach
incorporated in the QPM permits the description  of different 
processes, where excited states with intermediate energy are included:  
$\gamma$-decay \cite{Ponom2}, nucleon emission \cite{Giai}, ect.

The paper is organized as follows: 
In Section II some details of the $(\gamma, n)$ experiment are discussed and
the experimental results are presented.
A theoretical treatment of the $(\gamma, n)$ reaction is given in Section III.
In Subsection III.A we discuss the mechanism of the reaction while details
of the nuclear structure calculations in odd mass nuclei within the QPM
are given in Subsection III.B.  
The results of numeric calculations are discussed in Section IV in 
comparison with the experimental data.

\section{Experimental results.}

The measurements of the IR were performed on the electron accelerator 
microtron MT-25 of Flerov Laboratory of Nuclear Reactions, JINR.
The description of this microtron and its
main parameters were published earlier \cite{Gang}. The essential 
advantage of this accelerator is the small energy spread of the 
accelerated electrons (30 - 40 keV) at the high beam intensity 
(up to an average power 600 W). This allows measurement of the 
cross sections of the studied nuclides production at the strictly definite 
end-point energy.

The electron beam extracted from the accelerator chamber was focused 
by using the quadrupole lenses and  directing it on the stopping target, 
which consists of a  2 mm thick tungsten disk cooled by water. 
The electron spot on the target was about 5 mm. The mean current on 
the target, 20 $\mu$A, was measured continuously during the experiment. The 
energy of the electron beam (and the upper boundary of the 
bremsstrahlung) was changed by two methods:  the choice of the 
proper electron orbit and the tuning of the magnetic field.
 
The oxides of the studied elements -- BaO, Ce$_2$O$_3$, Nd$_2$O$_3$ 
and Sm$_2$O$_3$ were irradiated by the bremsstrahlung spectrum. 
They have a weight of $\sim$~100 mg, a surface area $\sim$~1~cm$^2$ and  
were enriched by the studied isotope up to 93\%.
The irradiation time was
10 min for each target. The irradiation flux was monitored by Cu activation
foils that were irradiated simultaneously with the samples. About one min
after the end of irradiation, the residual activity from the isotones
samples and Cu monitor, were measured separately by a calibrated
60~cm$^3$ Ge(Li) detector with an energy resolution
of $\Delta E =$ 2.5 keV for E$_{\gamma}$ = 1332 keV of $^{60}$Co.
The detector was operated in a lead shielding chamber lined with a 10 cm 
thick wall.
Measurements have been repeated to determine experimentally the half-life
of the radioactive nuclides under study. The spectra  were  processed
according to  the  ACTIV  code,  which permits separating $\gamma$-lines
with close energies in the complicated spectrum. Fig.~\ref{fig1} shows one
of the measured $\gamma$-spectrum obtained upon irradiating an enriched
$^{144}$Sm isotope by bremsstrahlung  with an endpoint energy of 25 MeV.
The $\gamma$-lines corresponding to $\beta - \gamma$ decay of unstable
$^{143}$Sm in the ground and the isomeric states into $^{143}$Pm are clearly
seen in the spectrum.
                                         
The IR is determined by the ratio of the measured nuclei yields in the 
isomeric and the ground state:	
\begin{equation}
\mbox{IR} = \frac{Y(\mbox{E}_{\gamma_{max}})_m}
{Y(\mbox{E}_{\gamma_{max}})_g}~,
\label{gan1}
\end{equation}
where the index $m$ and $g$ is attributed to the isomeric and ground
states, respectively; $Y$ is the yield of isomeric or ground state, and
E$_{\gamma_{max}}$ is the maximum endpoint energy.
The ratio of the yield is connected with the areas under the $\gamma$ peaks
corresponding to the isomeric and ground states ($S_{m}$ and $S_{g}$, 
respectively) by the equation
\begin{eqnarray}
\frac{Y(\mbox{E}_{\gamma_{max}})_m}
{Y(\mbox{E}_{\gamma_{max}})_g} &=& \left[
\frac{\lambda_g  f_m(t)}{\lambda_m  f_g(t)} ~
\left(
\frac {S_g \epsilon_m I_m (1+\alpha_g)}{S_m \epsilon_g I_g (1+\alpha_m)}
\right . \right .
\nonumber \\
&-& \left . \left .
p~\frac {\lambda_g}{\lambda_g - \lambda_m} \right)
+ p~\frac {\lambda_m}{\lambda_g - \lambda_m}\right]^{-1}~, 
\label{gan2}
\end{eqnarray}
where $\alpha$ is a conversion coefficient of $\gamma$-radiation,
$I$ is the probability of the $\gamma$-ray emission,
$\epsilon$ is the efficiency of the Ge(Li) detector for the corresponding
$\gamma$-line,
$\lambda$ is the decay constant,
$p$ is the probability of the transition from an isomeric
to a ground state,
and the time factor $f(t)$ that takes into account the accumulation and 
decay of the nuclei is given by
\begin{equation}
f(t_i,t_c,t_m)=(1-e^{-\lambda_g t_i})e^{-\lambda_g t_c}(1-e^{-\lambda_g t_m})~,
\label{gan3}
\end{equation}
where $t_{i}$, $t_{c}$, and $t_{m}$ are the times of irradiation, decay, and
measurement. Because the isomeric and ground states were obtained in the same
exposure run, and because the required quantities (yields or cross sections)
were measured for them under identical conditions, the errors associated
with the intensities of the bremsstrahlung flux and the geometry of 
exposures and measurements were eliminated in the determining IRs.

The produced in $(\gamma,n)$ reactions nuclei are radioactive both in the
ground and isomeric states (except $^{137}$Ba). Therefore it was possible
to determine the yields of studied nuclei in the ground and isomeric states
simultaneously by measuring the intensity of the emitted $\gamma$-radiation
using Eq.~(\ref{gan2}). For $^{137}$Ba the IR determination by  direct  method
becomes impossible for the lack of the appropriate $\gamma$-lines in the ground
state. Thus, the possibilities of isomer production of $^{137m}$Ba isomer
via $(\gamma,n)$ reactions were investigated at energies $E_{\gamma}$ =
10 - 25 MeV. The  yield  of  $^{137m}$Ba  was compared with the yield of
monitoring reaction. The cross section for the reaction 
$^{63}$Cu$(\gamma$,n)$^{62}$Cu measured in a beam of 
quasimonochromatic photons  
was taken for the reference \cite{Berm}. 
Reconstruction of reaction cross section,
employing an iterative procedure, was depicted in our previous paper
\cite {Belov}. Fig.~\ref{fig2} displays the cross section 
for $(\gamma,n)$ reactions
with production of the $^{137m}$Ba isomeric state. 
In order to determine IR, the $^{137m}$Ba cross section was subtracted 
from the total $^{137m+g}$Ba cross section \cite {Berm}. Thus, in 
the case of $^{137}$Ba, the IR was defined as a cross section ratio. 
The values of these IR are in accordance with the previous results 
\cite{Belov,Mazur1}. 
The IR dependence on the endpoint energy is also given in Fig.~\ref{fig2}.
Three main regions of IR behavior are recognized. 
The first region is from the neutron threshold to a maximum in the
photoabsorption cross section where the IR increases sharply.
In the second region (from 16 to 21~MeV) of the high-energy tail of the GDR, 
the IR changes slightly. 
Above 21~MeV where the GDR tail becomes very weak, the IR turns out to be
practically constant.
Since the spin and parity of a compound nucleus does not change (it is always
$1^{-}$ after dipole absorption), the observed increase in the IR is
due to the enlargement of the interval of excitation energies of a
residual nucleus in which photons that result in the 
population of the isomeric state occur. 

The IR at E$_{\gamma_{max}}$ = 25 MeV are presented in 
the Table~\ref{tab1}. 
The values of these IR are in accordance with 
the previous results \cite{Belov,Mazur2}, but they are more precise. 
We have also added the IR for $^{141}$Nd and $^{143}$Sm from the
$\beta$-decay of $^{141}$Pm and $^{143}$Eu, respectively. 
They are obtained from the analysis of the decay schemes of $^{141}$Pm 
and $^{143}$Eu \cite{Data}.

\section{Theoretical treatment of the isomer population in 
${ (\gamma, n)}$ reactions.}

The process of the isomer population in $(\gamma, n)$ reactions is rather 
complex
and may be split into a few stages. In the first stage, the GDR is
excited by bremsstrahlung radiation. In the second, it quickly decays 
into the compound nucleus. Next, the compound nucleus decays by emitting of 
neutrons, leaving the (A-1) nucleus with an excitation energy of a few MeV. 
In the last stage, the nucleus deexcites by direct or cascade 
$\gamma$-transitions into the isomeric or ground state.
Since the width of $\gamma$-decay strongly decreases with an increase
of the multipolarity, the decays are mainly due to
$E1$-, $M1$- or $E2$-transitions. 
The excited states of the nucleus at intermediate energies, via which
the isomers are populated, are called activated states (AS).
The specific properties of the AS are the following: First, due to
their total angular momentum and some components of their wave 
function, they tend to decay into the isomer than into the ground state.
Secondly, the AS should be excited themselves. The last depends
on the nuclear reaction considered. For example, in bremsstrahlung
experiments with the end-point energy of a few MeV, the AS state must
have a total angular momentum not very different from those of the 
ground state in order to make their electromagnetic excitation with low multipolarity
from the ground state possible. If $(\gamma, n)$ reaction is applied, the AS
should have the total spin close to 1 since emitted neutrons in the GDR decay
tend to take away a small angular momentum.
Thus, in principle, different AS may
be involved in the population of isomers in the different nuclear reactions. 

To some extent, the observed variations in the IR values can be associated
with the distinctions between the excitation energy and the angular momentum
distributions of the final nucleus prior to the photon direct or
cascade decay populating the isomeric and ground state.
In $(\gamma, n)$ reactions, the excitation energy and the angular 
momentum distributions are determined by
neutron emission from an excited nucleus produced as the result
of photon absorption. 
The angular momentum distribution for the reaction 
$(\gamma,n)$ on Ba, Ce, Nd, and Sm isotopes are similar. 
In this process, the neutron it emits from the state of an initial 
nucleus with 
spin and parity $1^{-}$ (dipole absorption) is followed by the population 
of the residual nucleus state with a spin $3/2^{+}$ (ground state)
or $11/2^{-}$ (isomeric state). When the isomer population is such that all 
the nuclei under study are characterized by equal spins at each steps of the 
reaction, the IR is determined by the excitation energy 
with which the nuclei 
rest after neutron decay. In the latter case, the energy distribution was
comparatively broad due to the bremsstrahlung spectrum of the $\gamma$
radiation. The width of this distribution is mainly defined by the  
photoabsorption Lorentzian curve which is very similar
to the studied isotones. The mean value of the excitation energy 
distribution, E$^*$, equals
\begin{equation}
{\rm E}^{*} = {\rm E}_{eff} - B_{n} - \varepsilon_{n}~,
\label{gang4}
\end{equation}
where E$_{eff}$ is the effective excitation energy, $B_{n}$ is the neutron
binding energy, and  $\varepsilon_{n}$ is the kinetic energy of the neutron
escaping from the compound nucleus. 
A peculiarity of the photonuclear reaction with a bremsstrahlung radiation
is related to the fact that the whole bremsstrahlung spectrum is
involved in the process of photoexcitation.
The energy centroid is determined by the ratio
\begin{equation}
{\rm E}_{eff} = \frac{\int_{{\rm E}_{th}}^{{\rm E}_{\gamma_{max}}}
{\rm E}_{\gamma}~ \sigma({\rm E}_{\gamma})~ 
N({\rm E}_{\gamma},{\rm E}_{\gamma_{max}})~d{\rm E}_{\gamma}}
{{\int_{{\rm E}_{th}}^{{\rm E}_{\gamma_{max}}}\sigma({\rm E}_{\gamma})~ 
N({\rm E}_{\gamma},{\rm E}_{\gamma_{max}})~
d{\rm E}_{\gamma}}}~,
\label{gan5}
\end{equation}
where ${\rm E}_{th}$ is a threshold energy, 
$\sigma({\rm E}_{\gamma})$ is the cross section for the absorption of a 
photon with energy E$_{\gamma}$ by a nucleus, and 
$N({\rm E}_{\gamma},{\rm E}_{\gamma_{max}})$ 
is the number of photons with energy ${\rm E}_{\gamma}$ in the
bremsstrahlung spectrum. 
The experimental kinetic energy spectra of the neutrons with a mean 
energy of 1 MeV were used. The maximum excitation energy of the residual 
nuclei after $(\gamma, n)$ reaction is presented in Table~\ref{tab1}. 

\subsection{Reaction mechanism.}

The main features of the $(\gamma, n)$ process are discussed 
in Ref. \cite{Bohr}.
The cross section for $(\gamma ,n)$ reaction with a population of
the isomeric (ground) state may be schematically written as:
\begin{equation}
\sigma _{iso(g.s.)}=\sigma_0 \cdot P_n^{lj} 
\cdot \frac{\Gamma_{iso(g.s.)}} {\Gamma_{tot}}~,  
\label{sigma1}
\end{equation}
where $\sigma_0$ is the photoabsorption cross section which describes the
GDR excitation with formation of a compound nucleus and
$\Gamma_{iso}$, $\Gamma_{g.s.}$, and $\Gamma_{tot}$ are
$\gamma$-widths for decay 
to isomeric, ground state and total decay width, respectively.
  
A relative probability $P_n^{lj}$ of emission of a
neutron with spin$\ j$ and angular momentum $l$ 
is given by the relation:
\begin{equation}
P_n^{lj} =\frac{\sigma_n^{lj}}
{\sum\limits_{l^{\prime }j^{\prime }} 
\sigma_n^{l^{\prime }j^{\prime }}}~.
\label{sigma}
\end{equation}
The quantity $\sigma _n^{lj}$ is the cross section of the
emission of a neutron with spin $j$ and angular momentum $l$ with formation
of a nucleus with momentum $J$ and projection $M$ and is equal to
\begin{equation}
\sigma_n^{lj}=\frac \pi {k^2}\frac{(2j+1)}2T_{lj}(\varepsilon )\left|
(jJ';0m\mid jJ';JM)\right| ^2,  \label{sigma2}
\end{equation}
here $(jJ';0m\mid jJ';JM)$ is a Clebsch-Gordan coefficient, $J'$ 
is the spin of the compound nucleus;  
neutron wave number is denoted by $k,$ the quantities $T_{lj}(\varepsilon)$ 
are the neutron penetrations. The energy $\varepsilon $ corresponds to
the average energy of the evaporated neutron from the compound nucleus.
The experimental values of the factors $T_{lj}(\varepsilon )$
at $\varepsilon = 1$~MeV are presented in Table~\ref{tab2}.
At this energy, the neutron could carry an angular moment 
$l=0,1,2$ and $3\cite{Mazur1}$. The probabilities for $l\geq 4$ are very
small and may be excluded from the consideration.
In the actual calculation the GDR energy centroid has been taken from
Ref.~\cite{Berm} and its width has been neglected. 

Combining Eqs. (\ref{sigma1}-\ref{sigma2})\cite{Optical}, 
the ratio $\sigma _{iso}/\sigma _{g.s.}$ 
for the ($\gamma ,n$) reaction reads
\begin{equation}
\frac{\sigma _{iso}}{\sigma _{g.s.}}=\frac{\sum\limits_{J_1^\pi
}(2J_1+1)T_{l_1j_1}(\varepsilon )\sum\limits_\vartheta \left( C_{J_1^\pi
}^\nu \right) ^2\Gamma _{J_1^\pi \rightarrow {iso}}^\vartheta }
{\sum\limits_{J_2^{\pi ^{^{\prime }}}}(2J_2+1)T_{l_2j_2}(\varepsilon
)\sum\limits_\vartheta \left( C_{J_2^{\pi ^{^{\prime }}}}^\nu \right)
^2\Gamma _{J_2^{\pi ^{^{\prime }}}\rightarrow {g.s.}}^\vartheta },  
\label{IR}
\end{equation}
where $j_{1(2)}$ and $l_{1(2)}$ are the moments of the emitted neutron
from the compound nucleus, $j_{1(2)}=\left| l_{1(2)}\pm 1/2\right| $
and $J_{1(2)}=$ $j_{1(2)}+1,$ the quantities $\left( C_J^\nu \right)
^2 $ are the spectroscopic factors for the $\vartheta $ state with momentum
and parity $J^\pi $ of the final nucleus; 
$\Gamma _{J^\pi \rightarrow{iso}}^\vartheta$ 
($\Gamma _{J^\pi \rightarrow{g.s.}}^\vartheta$)
are the partial widths for the $\gamma $-decay of the state 
$J^\pi $ to the isomeric (ground) state.
The value of 
$\Gamma ^\vartheta_J $ includes, in principle, 
the all unknown cascades which the intermediate state $J^\pi $ undergoes.

The partial widths $\Gamma $ are connected with the corresponding reduced
transition probabilities \cite{Bohr}:
\begin{equation}
\Gamma _{J^\pi \rightarrow I^{\pi ^{^{\prime }}}}({\cal X}\lambda )\sim
\mbox{E} ^{2\lambda +1}B({\cal X}\lambda ;J^\pi \rightarrow I^{\pi
^{^{\prime }}})~,
\end{equation}
here $B({\cal X}\lambda )$ are the reduced transition probabilities in 
$e^2fm^{2\lambda }$ for the electric and in $\mu _N^2fm^{2\lambda -2}$ for
the magnetic transitions; the quantity E is the $\gamma$-quanta
energy.

\subsection{Structure of excited states in odd nuclei.}

Nuclear structure calculations have been performed within the QPM (see,
e.g. Refs. \cite{Gale,Vdov1,sol1,sol2,sol3,Vdov2,Vor,Bert}).
This model has already been used in spectroscopic studies of odd nuclei 
over a wide mass region \cite{Gale,Vdov1}. 
The application of the model in the case of isomeric states 
is described in detail in Refs. \cite{Ponom1,Cosel,Carroll,Huber}. 
This model makes use of a separable form of the residual interaction which
enables the diagonalization of the model Hamiltonian in a large dimensional
configuration space. 

The following wave function describes the ground and excited
states of the odd nucleus with angular momentum $J$ and projection $M$ 
\begin{equation}
\Psi ^\vartheta _{JM}=C_J^\vartheta \left\{ \alpha _{JM}^{+}+\sum_{\lambda
\mu i}D_j^{\lambda i}(J\vartheta )\left[ \alpha _{jm}^{+}Q_{\lambda \mu
i}^{+}\right] _{JM}\right\} \Psi _{0}~.  \label{wave}
\end{equation}
The notation $\alpha _{jm}^{+}$ is the quasiparticle creation operator
with shell quantum numbers $j\equiv (n,l,j)$ and $m$; 
$Q_{\lambda \mu i}^{+}$ denotes the
phonon creation operator with the angular momentum $\lambda $, projection 
$\mu $ and RPA root number$\ i$; $\Psi _{0}$ is the ground state of
the neighboring even-even nucleus and $\vartheta $ stands for the number  
within a sequence of states of given $J^\pi$. Coefficients 
$C_J^\vartheta $ and $D_j^{\lambda i}$ are the quasiparticle and 
`quasiparticle$\otimes$phonon' amplitudes for the $\vartheta $ 
state. 

The  coefficients of the wave function (\ref{wave}), as well as the energy
of the excited states, are found by diagonalization of the model Hamiltonian 
within the approximation of the commutator linearization.
The components $\alpha _{jm}^{+}Q_{\lambda \mu i}^{+}$ of the wave function 
(\ref{wave}) may violate the Pauli principle. To solve this problem,
the exact commutation relations between quasiparticle and phonon operators 
are used \cite{Gale,Vdov1}. An efficient procedure for the approximate
treatment of the Pauli principle corrections has been proposed in Ref. 
\cite{Khuong}. It is applied to the present calculations.

It should be noted that the wave function (\ref{wave}) describes in detail
the distribution of the single-particle component while the distribution of the
`quasiparticle$\otimes $phonon' components could be more affected by including
`quasiparticle$\otimes $two-phonon' configurations \cite{Gale,Vdov1}. 
As demonstrated below in actual calculation, the transitions between 
quasiparticle components of the decaying intermediate and final states are 
mainly responsible for the population of the isomeric and ground states in 
$(\gamma, n)$ reaction.
For this reason, the wave function in the form of Eq.~(\ref{wave}) should be
considered as a suitable one to describe the process.     

Only the direct $\gamma$-decays of the intermediate states of the residual nuclei
into the ground and isomeric states have been accounted for in the
present studies. The lowest excited states of semi-magic even-even
nuclear core of isotopes under consideration have an excitation energy 
from 1.4 to 1.7~MeV. This means that only a few 
`quasiparticle$\otimes $three-phonon' and a very limited number of
`quasiparticle$\otimes $two-phonon' configurations are available 
up to an excitation energy of 5.4~MeV. Thus, a configuration space for 
the cascade decays is very small and the direct decays should play a
predominant role.

Woods-Saxon potential, $U(r)$, is used for the mean-field part of the QPM 
Hamiltonian with parameters from Refs. \cite{Chep,Tak}. 
The corresponding single-particle spectra can be found in Ref. \cite{Gale}.
The residual particle-hole interaction is taken of separable form in
coordinate space with radial form factor as $f(r)=dU/dr$.
The strength of the residual interaction is adjusted to reproduce in 
one-phonon approximation the experimental energies and transition
probabilities\cite{Endt} of the lowest collective 
levels in the neighboring even-even nucleus for each $J^{\pi}$.
The phonon basis is obtained by solving the QRPA equations. The coupling
matrix elements between quasiparticle and 'quasiparticle$\otimes$phonon'
configurations are calculated  using the internal fermion structure
of the phonons and the model Hamiltonian
 with all parameters fixed from the calculations
in the even-even core. 

In the calculations  presented below we have used 
$\lambda^{\pi} = 1^{\pm }, 2^{+}, 3^{-},4^{+}$ and $5^{-}$
phonons in the wave function (\ref{wave}). Several roots for each
multipolarity are taken into account. The maximum energy of
'quasiparticle$\otimes$phonon' configurations included in (\ref{wave})
equals 12~MeV.

The used effective charges are:
for $E1$ transitions $e_{{\rm eff}}^p=(N/A)\,e$ and 
$e_{{\rm eff}}^n=-(Z/A)\,e$ to separate a center of mass motion;
``free'' values for $E2$ transitions $e^{n}=0$ and $e^{p}=1$ because
our single-particle basis is rather complete; and
for $M1$ transitions the effective spin g$^s$ factor is 
g$_{eff}^s$=(0.8)g$_{free}^s$. The last value often has been used in
the past in QPM calculation of magnetic moments \cite{Vdo90}, low energy
magnetic transitions \cite{low} and excitation probability of M1 and M2 
resonances \cite{M1M2} in medium and heavy nuclei. The reduction factor 0.8
provides the best agreement with experimental data available.
The numerical calculations presented in this article have been performed with
the code PHOQUS\cite{Khuong}.

\section{Theoretical analysis of the experimental data.}

The IR is studied along the isotopic chain N=81 including nuclei $^{137}$Ba,
$^{139}$Ce, $^{141}$Nd and $^{143}$Sm. 
First,  the properties of the $(11/2^{-})$ isomeric state
are reproduced well in calculation.
Its excitation energy is known to be  approximately 
the same for all nuclei under consideration (see experimental data in 
second column of Table~\ref{tab3}).  
The calculated energy of the isomeric state (third column of Table~\ref{tab3})
is in agreement with the data. Its structure is dominated by the
single-particle component, but the coupling with the quadrupole and
octupole vibrations of the core is also important. The calculated structure of
the isomeric state is also in agreement with the recent 
measurements \cite{Data}.

In addition, the spectrum of the excited states has been calculated with 
$J^\pi =1/2^{\pm },3/2^{\pm},5/2^{\pm },7/2^{\pm },9/2^{+}$ and $11/2^{-}$
up to the excitation energy of 6.5 MeV. 
The states with lower values of angular momentum are responsible for 
population of the ground state in $(\gamma ,n)$ reaction while among other
states we are looking for the AS. 
The spectra of states with $J > 11/2$ have not been calculated because they
cannot be populated in the neutron decay of the GDR. 
The corresponding partial widths $\Gamma
_{J_i^{}\rightarrow J_f}$ and the main components contributing in
the structure of the excited states are given in Table~\ref{tab4} 
for $^{139}$Ce.

The population of the isomeric state $(11/2^{-})$ in $^{139}$Ce is mainly
due to the $E1$ transitions from the $J^\pi =9/2^{+}$ states. The
distribution of the single-particle component $9/2^{+}$ of the wave function 
(\ref{wave}) up to 6.5 MeV is shown in Fig.~\ref{fig3}(b). 
A visible fraction of $9/2^{+}$ single-particle component 
(14\% of the whole strength) is spaced in this energy region. 
There are two excited states at $4.18$ and $6.15$
MeV, where 4.7\% and 5.2\% of the single-particle strength is concentrated.
The main components in the structure of the excited states are due to the
coupling of the single-particle mode with the surface vibrations of the even
core. The dependence of the quantity $C^2\Gamma _{J^\pi \rightarrow iso}$
on the excitation energy is shown in Fig.~\ref{fig3}(a). The
contribution of the $E1$ transitions dominates, but also $E2$ transitions
are important. The comparison of Fig.~\ref{fig3}(a) and 
Fig.~\ref{fig3}(b) reveals a bright correlation between the distribution 
of $C^2\Gamma _{9/2^{+}\rightarrow iso}$ $(E1)$ and those of the 
single-particle
component $9/2^{+}$. The higher concentration of the single-particle component
leads to larger $E1$ transitions into the isomeric state.

The $7/2^{-}$ states populate the isomeric state by $E2$ transitions. The
distribution of the single-particle component $7/2^{-}$ of the wave function
(\ref{wave}) is shown in Fig.~\ref{fig3}(c). 
There are two main states at $3.22$ 
MeV and $4.88$ MeV where 0.6\% and 0.4\% of the single-particle strength
is concentrated. By comparing Fig.~\ref{fig3}(a) and Fig.~\ref{fig3}(c), 
it can be  concluded that there is also a correlation between the 
distribution of the single-particle component 
$7/2^{-}$ and the quantity $C^2\Gamma _{7/2\rightarrow iso}(E2)$: large
single-particle component in the wave function (\ref{wave}) leads to larger 
$E2$ transitions. Taking into account the foregoing properties of $E1$ and 
$E2 $ transitions one concludes that the structure of the quantity $C^2\Gamma
_{J^\pi \rightarrow iso}$ is completely determined by the contribution of
the single-particle component in the wave function (\ref{wave}).

The energy dependence of the quantity 
$C^2\Gamma _{J^\pi \rightarrow g.s.}$
in $^{139}$Ce and the contribution of $E1$, $E2$ and $M1$ transitions are 
shown in Fig.~\ref{fig4}. 
These transitions are due to the deexcitation of 1/2$^{\pm },$ 
3/2$^{\pm },$ 5/2$^{+}$ and 7/2$^{+}$ states. 
The large $E1$ transitions to the ground
state are concentrated at higher excitation energies. A state with $J^\pi
=1/2^{-}$ at $5.35$ MeV gives the main part of the $E1$ transitions to the
ground state in $^{139}$Ce. In the structure of the state, the
single-particle component is 6\%, while the $\left( 2d_{5/2}\otimes
3_1^{-}\right) $component contributes 61.2\%.

The $M1$ transitions to the ground state are predominantly due to the
deexcitation of the 5/2$^{+}$  states. The excitation energy of the first 
5/2$^{+\text{ }}$state is $1.49$ MeV. The single-particle component 
dominates in
the structure (68\%) and it determines the largest value of $C^2\Gamma (M1)$
in the domain up to 6.5 MeV.

The $E2$ transitions to the ground state arise from the deexcitation of the
states $1/2^{+},3/2^{+},$ $5/2^{+}$ and $7/2^{+}$. 
The state $1/2^{+}($2.93~MeV) has
the largest income in the population of the ground state via $E2$
transition. The single-particle component in the structure of the state is
5.5\%. The E2 transition from the latter to the ground state presents
around 6.5\% of the sum of $\sigma _{g.s.}$ in Eq. (\ref{IR}).

The distribution of the single-particle configuration over the states with
$J^{\pi}= 1/2^{-}$, 3/2$^{-},
$ 5/2$^{+}$ and 7/2$^{+}$ in $^{139}$Ce up to 6.5 MeV 
is shown in Fig.~\ref{fig4}(b,c,d,e). 
The comparison of Fig.~\ref{fig4}(a) and 
Fig.~\ref{fig4}(b,c,d,e) reveals again a correlation between the 
quantity $C^2\Gamma _{J^\pi \rightarrow
g.s.}$ and those of the single-particle strength. The energy dependence of 
$C^2\Gamma _{J^\pi \rightarrow g.s.}$ is strongly connected to the amplitude
value  of the single-particle component of the wave function (\ref
{wave}). The structure of $1/2^{+}$ and $3/2^{+}$ states and the corresponding
value of $\Gamma _{J^\pi \rightarrow g.s.}$ are given in Table~\ref{tab4}. The
population of the ground up to 4.8 MeV is by E2 and M1 transitions. Only a
few states perceptibly influence  the value of $C^2\Gamma _{J^\pi \rightarrow
g.s.}$.

The same type of calculation have been performed for the other isotones. 
They are not discussed here in detail because the results and conclusions are 
very similar to the ones already presented for $^{139}$Ce.
The conclusion for all isotones under consideration is
that the population of the isomeric state as well as those for the
ground state depends on the distribution of the single-particle
component of the wave function ($\ref{wave}$) over the AS.

It follows from the presented figures that the IR defined in Eq. (\ref{IR})
depends strongly on the maximum excitation energy of the final nucleus. 
The number of excited states at intermediate energies involved in the
process varies when the excitation energy is changed and the number of large 
$E1$, $E2$ and $M1$ transitions populating isomeric and ground state is also
changed. The available experimental information obtained in $(\gamma ,n)$
reaction on the IR for N=81 nuclei is given in Table~\ref{tab1}. 
It is completed by the IR in two heaviest isotones known from 
$\beta$ decay data \cite{Data}.
Thus, in the case of $^{141}$Nd and $^{143}$Sm, two values of the IR are known
corresponding to different maximum excitation energies, E$^{*}$. 
The calculated values of the
IR of Eq. (\ref{IR}) are given in  Table~\ref{tab1} for comparison. 
The truncation of the AS included in the calculation
 by maximum excitation energy
is made in accordance with experimental conditions. 
The calculated values agree very well with
the measured ones. The energy dependence of the IR in the case of $^{141}$Nd
and $^{143}$Sm is reproduced. In the case of $^{143}$Sm up to
excitation energy 3.5 MeV there are two states - $9/2^{+}$ at 3.48 MeV
and $7/2^{-}$ at 3.46 MeV. These states are connected with the isomer
by E1 and E2 transitions. The increase of the excitation energy up to 4.72 
MeV does not influence the value of $C^2\Gamma _{J^\pi \rightarrow 
iso}$ while the value of $C^2\Gamma _{J^\pi \rightarrow g.s.}$ is
changed due to the $E2$ and $M1$ transitions connected the $1/2^{+},5/2^{+}$
and $7/2^{+}$ states with the ground state. The situation is similar in 
$^{141}$Nd.

\section{Conclusions.}

The IR in N = 81 and Z = 56 -- 62 nuclei 
obtained in ($\gamma$,n) reaction on the 
beam of bremsstrahlung were measured. The large difference of the IR for the 
studied nuclei was observed in spite of the similar properties of the isomeric 
states.
The measured values are completed by available experimental information 
obtained from $\beta^-$ decay. The new set of data reveals a dependence of
the IR on a maximum excitation energy of residual nuclei
reached in ($\gamma$,n) reaction and $\beta^-$ decay.

The calculation of the IR in the framework of QPM is done. It is shown that
the value of the IR depends on the contribution of the corresponding
single-particle component in the structure of the wave function of the
excited state. The IR as a function of the excitation energy of a
residual nucleus is determined by the fluctuation in the distribution
of the single-particle components. The maximum energy of the intermediate 
states populated in ($\gamma$,n) reaction is responsible for the 
truncation of the states involved in the forthcoming decay to the ground
and isomeric states.
That is why the IR in $^{137}$Ba and $^{139}$Ce with somewhat higher
values of the maximum excitation energy is different from the ones in 
$^{141}$Nd and $^{143}$Sm. Our calculation reproduces rather well the IR
dependence on A as well as the absolute values of the IR in all
isotones under consideration.

The present work is partly supported by the Bulgarian Science Foundation
(contract Ph. 626) and Plovdiv University (contract PU F--23). 
V. Yu. P. acknowledges the support from the 
Research Council of the University of Gent and a NATO fellowship.

Finally, we would like to thank Lynn Leonard for her careful reading 
of this paper.

\begin{figure}[htbp]
\caption{ Gamma-ray spectrum measured after activation of
$^{144}$Sm  target by bremsstrahlung radiation with endpoint energy
$E_ {\gamma_{max}}$= 25 MeV. The $\gamma$-lines used in the data analysis are
indicated for the isomeric and ground states.}
\label{fig1}
\end{figure}

\begin{figure}[htbp]
\caption{ Measured cross section for the reaction
$^{138}$Ba$(\gamma, n)^{137m}$Ba as a function of bremsstrahlung
end-point energy. The isomeric ratio is plotted with the right-hand side axis.}
\label{fig2}
\end{figure}

\begin{figure}[htbp]
\caption{(a) Decay probability, $C^2 \Gamma$, for population of the isomeric
state in $^{139}$Ce from excited states at intermediate energies; 
(b)-(c) contribution of one-quasiparticle configuration, $C^2$, 
to wave functions of
decaying states with different spin and parity. Multipolarity of each
transition is indicated on top of lines in (a). $J^{\pi}$ of decaying
states is specified in (b) and (c).}
\label{fig3}
\end{figure}

\begin{figure}[htbp]
\caption{The same as in Fig.~\protect{\ref{fig3}} for population of
the ground state in $^{139}$Ce.}
\label{fig4}
\end{figure}

\begin{table}
\caption{The calculated (QPM) isomeric ratio for $(\gamma ,n)$ reaction and 
$\beta$ decay in comparison with the experimental results (Exp).
E$^{*}$ is the maximum energy of intermediate states reached in every
reaction. 
\label{tab1}} 
\begin{tabular}{lllll} 
Reaction & \multicolumn{2}{c}{E$^{*},$ MeV} & \multicolumn{2}{c}{Isomeric
ratio} \\ 
& Exp & QPM & Exp & QPM \\ \hline 
$^{138}$Ba$(\gamma ,n)^{137}$Ba & 5.4 & 5.4 & 0.12(1) & 0.10 \\ 
$^{140}$Ce$(\gamma ,n)^{139}$Ce & 4.8 & 4.8 & 0.14(1) & 0.11 \\ 
$^{142}$Nd$(\gamma ,n)^{141}$Nd & 4.2 & 4.2 & 0.06(1) & 0.05 \\ 
$^{144}$Sm$(\gamma ,n)^{143}$Sm & 3.5 & 3.5 & 0.047 & 0.051 \\ 
$^{143}$Eu$\stackrel{\beta }{\rightarrow }$ $^{143}$Sm & 4.56 & 4.6 & 0.009 & 
0.007 \\ 
$^{141}$Pm$\stackrel{\beta }{\rightarrow }$ $^{141}$Nd & 5.1 & 5.1 & 0.01 & 
0.01 \\ 
\end{tabular}
\end{table}

\begin{table}
\caption{The experimental values of $T_{lj}(\varepsilon )$ for different
angular momentum $l$ of the emitted neutron with the energy 
$\varepsilon = 1$~MeV.
\label{tab2}}
\begin{tabular}{lllllllll}
&  & $l=0$ &  & $l=1$ &  & $l=2$ &  & $l=3$ \\ \hline
$T$  &  & 0.55 &  & 0.33 &  & 0.05 &  & 0.006 \\ 
\end{tabular}
\end{table}

%\end{multicols}

\widetext

\begin{table}
\caption{The calculated energies (QPM) of the isomeric state $h_{11/2}$
in $^{137}$Ba, $^{139}$Ce, $^{141}$Nd and $^{143}$Sm in comparison with
the experimental values (Exp) from \protect\cite{Data}. 
The main components of the isomeric wave function are presented in 
last two columns.
\label{tab3}}
\begin{tabular}{lllll} 
& \multicolumn{2}{c}{E, MeV} & \hspace*{7mm}$\alpha ^{+}$ 
& \hspace*{23mm}$\alpha ^{+}Q^{+}$ \\ 
& Exp & QPM &  &  \\ \hline
$^{137}$Ba & 0.662 & 0.660 & $1h_{11/2}(83.9\%)$ & 1h$_{11/2}\otimes
2_1^{+}$(7.2\%)+2d$_{5/2}\otimes 3_1^{-}(1.5\%)$ \\ 
$^{139}$Ce & 0.754 & 0.725 & $1h_{11/2}(84.9\%)$ & 1h$_{11/2}\otimes
2_1^{+}$(8.1\%)+2d$_{5/2}\otimes 3_1^{-}(1.4\%)$ \\ 
$^{141}$Nd & 0.757 & 0.795 & $1h_{11/2}(86.2\%)$ & 1h$_{11/2}\otimes
2_1^{+}$(5.6\%)+2d$_{5/2}\otimes 3_1^{-}(1.9\%)$ \\ 
$^{143}$Sm & 0.754 & 0.786 & $1h_{11/2}(84.6\%)$ & 1h$_{11/2}\otimes
2_1^{+}$(6.2\%)+2d$_{5/2}\otimes 3_1^{-}(2.5\%)$ \\ 
\end{tabular}
\end{table}

\begin{table}
\caption{Some selected excited states in $^{139}$Ce obtained in 
QPM calculations in
the energy range up to 6.5~MeV with the largest values of the partial widths, 
$\Gamma _{J_i\rightarrow J_f}$, for the direct decay into the ground and 
isomeric states. The quasiparticle ($\alpha^+$) and the main 
`quasiparticle$\otimes $phonon' $\left( \alpha ^{+}Q^{+}\right) $ 
components to their wave function are given in last two columns.
\label{tab4}}
\begin{tabular}{lllllll}  
$J_i$& \hspace*{1.5mm}E, & Tran. & $J_f$  
& $\Gamma _{J_i\rightarrow J_f},$ & 
$\hspace*{5mm}\alpha^{+}$ & \hspace*{11mm}$\alpha ^{+}Q^{+}$ \\ 
& MeV & ${\cal X}\lambda $ &
& \hspace*{2mm}eV &  &  \\ \hline
1/2$^{+}$ & 2.93 & E2 & 3/2$^{+}(g.s.)$ & 0.046 & 3s$_{1/2}(5.5\%)$ & 
2d$_{3/2}\otimes 2_1^{+}(91.4\%)$ \\ 
1/2$^{+}$ & 4.40 & E2 & 3/2$^{+}(g.s.)$ & 0.259 & 3s$_{1/2}(0.4\%)$ & 
2d$_{3/2}\otimes 2_4^{+}(90.9\%)$ \\ 
1/2$^{+}$ & 4.80 & E2 & 3/2$^{+}(g.s.)$ & 0.017 & 3s$_{1/2}(3.8\%)$ & 
2d$_{5/2}\otimes 2_1^{+}(79.8\%)$ \\ 
1/2$^{+}$ & 5.13 & E2 & 3/2$^{+}(g.s.)$ & 0.008 & 3s$_{1/2}(1.3\%)$ & 
1g$_{7/2}\otimes 4_1^{+}(93\%)$ \\ 
1/2$^{-}$ & 5.35 & E1 & 3/2$^{+}(g.s.)$ & 2.420 & 2p$_{1/2}(6.0\%)$ & 
2d$_{5/2}\otimes 3_1^{-}(61.2\%)$ \\ 
1/2$^{-}$ & 5.64 & E1 & 3/2$^{+}(g.s.)$ & 0.142 & 2p$_{1/2}(0.3\%)$ & 
1g$_{7/2}\otimes 3_1^{-}(63.9\%)$ \\ 
3/2$^{+}$ & 2.82 & E2 & 3/2$^{+}(g.s.)$ & 0.045 & 2d$_{3/2}(1.2\%)$ & 
2d$_{3/2}\otimes 2_1^{+}(86.1\%)$ \\ 
3/2$^{+}$ & 3.08 & M1 & 3/2$^{+}(g.s.)$ & 0.002 & 2d$_{3/2}(3.7\%)$ & 
3s$_{1/2}\otimes 2_1^{+}(84.4\%)$ \\ 
3/2$^{+}$ & 4.40 & E2 & 3/2$^{+}(g.s.)$ & 0.270 & 2d$_{3/2}(0.3\%)$ & 
2d$_{3/2}\otimes 2_4^{+}(96.2\%)$ \\ 
3/2$^{-}$ & 3.46 & E1 & 3/2$^{+}(g.s.)$ & 0.013 & 2p$_{3/2}(1.1\%)$ & 
1h$_{11/2}\otimes 4_1^{+}(93\%)$ \\ 
3/2$^{-}$ & 5.09 & E1 & 3/2$^{+}(g.s.)$ & 0.056 & 2p$_{3/2}(1.5\%)$ & 
1h$_{11/2}\otimes 4_4^{+}(95.4\%)$ \\ 
3/2$^{-}$ & 5.53 & E1 & 3/2$^{+}(g.s.)$ & 0.030 & 2p$_{3/2}(0.7\%)$ & 
2$d_{5/2}\otimes 3_1^{-}(84.5\%)$ \\ 
5/2$^{+}$ & 1.49 & M1 & 3/2$^{+}(g.s.)$ & 0.009 & 2p$_{3/2}(68\%)$ & 
3s$_{1/2}\otimes 2_1^{+}(8\%)$ \\ 
5/2$^{+}$ & 2.78 & E2 & 3/2$^{+}(g.s.)$ & 0.049 & 2d$_{5/2}(0.7\%)$ & 
2d$_{3/2}\otimes 2_1^{+}(93.2\%)$ \\ 
5/2$^{+}$ & 3.21 & M1 & 3/2$^{+}(g.s.)$ & 0.008 & 2d$_{5/2}(5.3\%)$ & 
2d$_{3/2}\otimes 4_1^{+}(57.1\%)$ \\ 
5/2$^{+}$ & 4.32 & E2 & 3/2$^{+}(g.s.)$ & 0.047 & 2d$_{5/2}(2.2\%)$ & 
1h$_{11/2}\otimes 3_1^{-}(62.4\%)$ \\ 
5/2$^{+}$ & 4.4 & E2 & 3/2$^{+}(g.s.)$ & 0.226 & 2d$_{5/2}(0.8\%)$ & 
2d$_{3/2}\otimes 2_4^{+}(78.6\%)$ \\ 
5/2$^{+}$ & 4.9 & M1 & 3/2$^{+}(g.s.)$ & 0.016 & 2d$_{5/2}(2.9\%)$ & 
2d$_{5/2}\otimes 4_1^{+}(44\%)$ \\ 
5/2$^{+}$ & 5.01 & M1 & 3/2$^{+}(g.s.)$ & 0.011 & 2d$_{5/2}(2.4\%)$ & 
2d$_{5/2}\otimes 4_1^{+}(41.8\%)$ \\ 
5/2$^{+}$ & 5.13 & M1 & 3/2$^{+}(g.s.)$ & 0.028 & 2d$_{5/2}(2.5\%)$ & 
1g$_{7/2}\otimes 4_1^{+}(33.7\%)$ \\ 
5/2$^{+}$ & 6.29 & M1 & 3/2$^{+}(g.s.)$ & 0.018 & 2d$_{5/2}(1.3\%)$ & 
2d$_{5/2}\otimes 2_4^{+}(78\%)$ \\ 
7/2$^{+}$ & 2.02 & E2 & 3/2$^{+}(g.s.)$ & 0.006 & 1g$_{7/2}(59.9\%)$ & 
2d$_{3/2}\otimes 2_1^{+}(28.8\%)$ \\ 
7/2$^{+}$ & 2.98 & E2 & 3/2$^{+}(g.s.)$ & 0.03 & 1g$_{7/2}(10.6\%)$ & 
2d$_{3/2}\otimes 2_1^{+}(55.6\%)$ \\ 
7/2$^{+}$ & 3.17 & E2 & 3/2$^{+}(g.s.)$ & 0.005 & 1g$_{7/2}(6.7\%)$ & 
2d$_{3/2}\otimes 4_1^{+}(68\%)$ \\ 
7/2$^{+}$ & 3.38 & E2 & 3/2$^{+}(g.s.)$ & 0.002 & 1g$_{7/2}(6.4\%)$ & 
3s$_{1/2}\otimes 4_1^{+}(71.5\%)$ \\ 
7/2$^{+}$ & 4.43 & E2 & 3/2$^{+}(g.s.)$ & 0.271 & 1g$_{7/2}(1.4\%)$ & 
2d$_{3/2}\otimes 2_4^{+}(90.6\%)$ \\ 
7/2$^{+}$ & 4.85 & E2 & 3/2$^{+}(g.s.)$ & 0.008 & 1g$_{7/2}(3\%)$ & 
1g$_{7/2}\otimes 2_1^{+}(56.1\%)$ \\[1.5mm] 
7/2$^{-}$ & 3.22 & E2 & 11/2$^{-}(iso)$ & 0.045 & 1f$_{7/2}(0.6\%)$ & 
1h$_{11/2}\otimes 2_1^{+}(98.5\%)$ \\ 
7/2$^{-}$ & 4.88 & E2 & 11/2$^{-}(iso)$ & 0.232 & 1f$_{7/2}(0.4\%)$ & 
1h$_{11/2}\otimes 2_4^{+}(98.8\%)$ \\ 
9/2$^{+}$ & 4.18 & E1 & 11/2$^{-}(iso)$ & 0.793 & 1g$_{9/2}(4.7\%)$ & 
1h$_{11/2}\otimes 3_1^{-}(86.3\%)$ \\ 
9/2$^{+}$ & 4.62 & E1 & 11/2$^{-}(iso)$ & 0.417 & 1g$_{9/2}(1.7\%)$ & 
2d$_{5/2}\otimes 2_1^{+}(84.9\%)$ \\ 
9/2$^{+}$ & 5.03 & E1 & 11/2$^{-}(iso)$ & 0.188 & 1g$_{9/2}(0.6\%)$ & 
2d$_{5/2}\otimes 4_1^{+}(52.3\%)$ \\ 
9/2$^{+}$ & 6.15 & E1 & 11/2$^{-}(iso)$ & 3.07 & 1g$_{9/2}(5.2\%)$ & 
1h$_{11/2}\otimes 3_3^{-}(58.4\%)$ \\ 
9/2$^{+}$ & 6.30 & E1 & 11/2$^{-}(iso)$ & 0.326 & 1g$_{9/2}(0.5\%)$ & 
2d$_{5/2}\otimes 2_4^{+}(61\%)$ \\ 
\end{tabular}
\end{table}

\narrowtext


\begin{references}
\bibitem[*]{byline1}
Present address: Vakgroep Subatomaire en Stralingsfysica,
Universiteit Gent,
Proeftuinstraat 86,
9000 Gent, Belgium

\bibitem{Gang}  Yu. P. Gangrsky, A. P. Tonchev, and N. P. Balabanov, 
Fiz. Elem. Chastits At. Yadra {\bf 27}, 1043 (1996).

\bibitem{Belov}  A. G. Belov, Yu. P. Gangrsky, A. P. Tonchev, and
N. P. Balabanov, Yad. Phys. {\bf 59}, 585 (1996).

\bibitem{Mazur1}  V. M. Mazur, V. A. Jeltonojskij, and Z. M. Bigan, Yad. Phys. 
{\bf 58}, 970 (1995).

\bibitem{Mazur2}  V. M. Mazur, Z. M. Bigan, and I. V. Sokolyuk, Laser Phys. 
{\bf 5}, 273 (1995); V. M. Mazur, I. V. Sokolyuk, and Z. M. Bigan, Yad. Phys. 
{\bf 54}, 895 (1991).

\bibitem{Duben}  A. P. Dubenskiy, V. P. Dubenskiy, A. A. Bojkova, and L.
Malov, Bull. Acad. Sci. USSR, Ser. Phys. {\bf 57}, 90 (1993).

\bibitem{Ponom1}  V. Ponomarev, A. P. Dubenskiy, V. P. Dubenskiy and E. A.
Boykova, J. Phys. G: Nucl. Part. Phys. {\bf 16}, 1727 (1990).

\bibitem{Cosel}  P. von Neumann-Cosel, V. Yu. Ponomarev, A. Richter,
and C. Spieler, Z. Phys. A {\bf 350}, 303 (1995).

\bibitem{Carroll}  J. J. Carrol, C. B. Collins, K. Heyde, M. Huber, P. von
Neumann-Cosel, V. Yu. Ponomarev, D. G. Richmond, A. Richter, C. Schlegel, T.
W. Sinor, and K. N. Taylor, Phys. Rev. C {\bf 48}, 2238 (1993).

\bibitem{Huber}  M. Huber, P. von Neumann-Cosel, A. Richter, C. Schlegel, R.
Schulz, J.J. Carroll, K. N. Taylor, D. G. Richmond, T. W. Sinor, C. B.
Collins, and V. Yu. Ponomarev, Nucl. Phys. {\bf A559}, 253 (1993).

\bibitem{Tonch}  A. P. Tonchev, Yu. P. Gangrsky, A. G. Belov, and V. E.
Zhuchko, Phys. Rev. C {\bf 58}, 2851 (1998).

\bibitem{Arifov}  L. Ya. Arifov, B. S. Mazitov, and V. G. Ulanov, Yad. Phys. 
{\bf 34}, 1028 (1981).

\bibitem{Vanden}  R. Vandenbosch and J. R. Huizenga, Phys. Rev. {\bf 120},
1305, 1313 (1960).

\bibitem{Ignat}  A. V. Ignatjuk, G. N. Smirenkin, and A. S. Tishin, 
Yad. Phys. {\bf  21}, 485 (1975).

\bibitem{Gale}  S. Gales, Ch. Stoyanov, and A. I. Vdovin, Phys. Rep. 
{\bf 166}, 125 (1988).

\bibitem{Vdov1}  A. I. Vdovin, V. V. Voronov, V. G. Soloviev, and Ch.
Stoyanov, Part. Nucl. {\bf 16}, 245 (1985).

\bibitem{Ponom2}  V. Yu. Ponomarev, V. G. Soloviev, Ch. Stoyanov, 
and A. I. Vdovin, Phys. Lett. {\bf 183B, }237 (1987).

\bibitem{Giai}  Nguen Van Giai, Ch. Stoyanov, V. V. Voronov, and S. Fortier,
Phys. Rev. C {\bf 53}, 730 (1996).

\bibitem{Berm}  B. L. Berman, S. C. Fultz, J. T. Caldwell, M. A. Kelly,
and S. S. Dietrich, Phys. Rev. C {\bf 2}, 2318 (1970).

\bibitem{Data}  T. W. Burrows, Nuclear Data Sheets {\bf 57}, 399 (1989);\\
J. K. Tuli, Nuclear Data Sheets {\bf 72}, 369 (1994),
{\bf 45}, 106 (1985), {\bf 48}, 812 (1986).

\bibitem{Bohr} A. Bohr and B. Mottelson, {it Nuclear structure Volume 1
Single-particle motion}, Nordita, Copenhagen (1969).

\bibitem{Optical}  G. I. Marchuk and V. E. Kolesov, {\it Application of
numerical methods for calculation of neutron cross-sections, }Atomizdat,
Moskva (1970).

\bibitem{sol1}  V. G. Soloviev, {\it Theory of complex nuclei}
(Oxford: Pergamon Press, 1976 ).

\bibitem{sol2}  V. G. Soloviev, {\it Theory of atomic nuclei : Quasiparticles
and Phonons} (Institute of Physics Publishing, Bristol and Philadelphia, 1992).

\bibitem{sol3}  V. G. Soloviev, Ch. Stoyanov, and A. I. Vdovin, Nucl. Phys. 
{\bf A342}, 261 (1980).

\bibitem{Vdov2}  A. I. Vdovin and V. G. Soloviev, Part. Nucl. {\bf 14},
237 (1983).

\bibitem{Vor}  V. V. Voronov and V. G. Soloviev, {\it Part. and Nucl.} 
{\bf 14}, 1380 (1983).

\bibitem{Bert} C.A. Bertulani and V.Yu. Ponomarev, Phys. Rep.  
{\bf 321}, 139 (1999).

\bibitem{low} 
T.K. Dinh, M. Grinberg. and Ch. Stoyanov, J. Phys. G {\bf 18} 329
(1992); \\
M. Grinberg, Ch. Stoyanov, and N. Tsoneva,
Fiz. Elem. Chastits At. Yadra {\bf 29}, 1456 (1998).

\bibitem{Vdo90}
A.I. Vdovin, R.R. Safarov, and V.Yu. Ponomarev,   
Bull. Acad. Sci. USSR, Phys. Ser. {\bf 54(9)}, 149 (1990).

\bibitem{M1M2}
V.Yu. Ponomarev, V.G. Soloviev, Ch. Stoyanov, and A.I. Vdovin, 
Nucl. Phys. {\bf A323}, 446 (1979). \\
A.I. Vdovin, V.V. Voronov, V.Yu. Ponomarev, and Ch. Stoyanov, 
Sov. Journ. of Nucl. Phys. {\bf 30}, 479 (1979).

\bibitem{Khuong}  Ch. Stoyanov and C. Q. Khuong, JINR Dubna Report
P-4-81-234, 1981 (unpublished).

\bibitem{Chep}  V. A. Chepurnov, Sov. J. Nucl. Phys. {\bf 6}, 955 (1967).

\bibitem{Tak}  K. Takeuchi and P. A. Moldauer, Phys. Lett. {\bf 28B}, 384
(1969).

\bibitem{Endt}  P. M. Endt, At. Data Nucl. Data Tables {\bf 42}, 1 (1989);
S. Raman, W. C. Nestor, Jr., S. Kahane, and K. H.Bhatt, At. Data Nucl. Data
Tables {\bf 42}, 1 (1989).

\end{references}
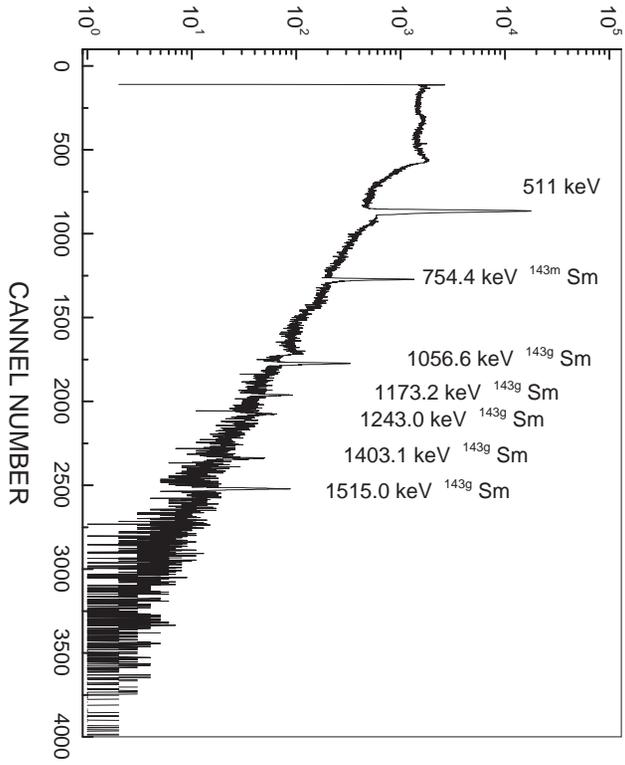
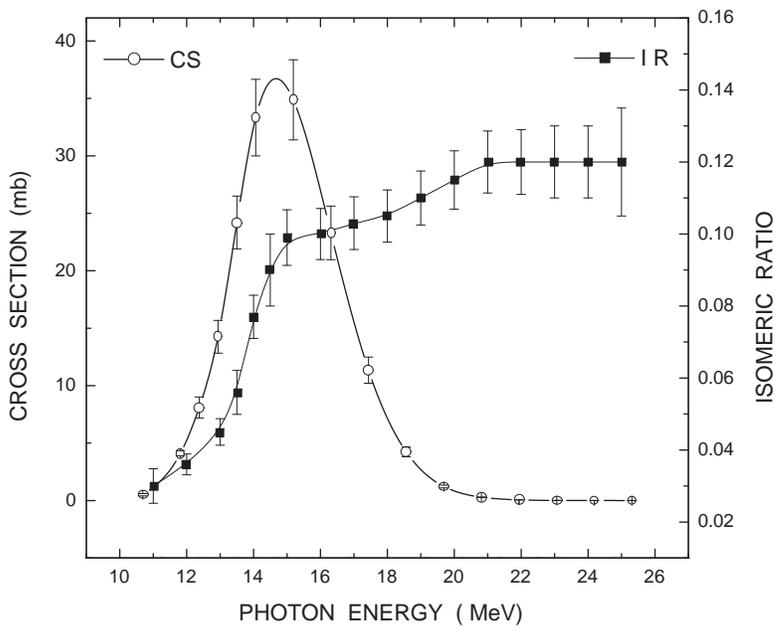
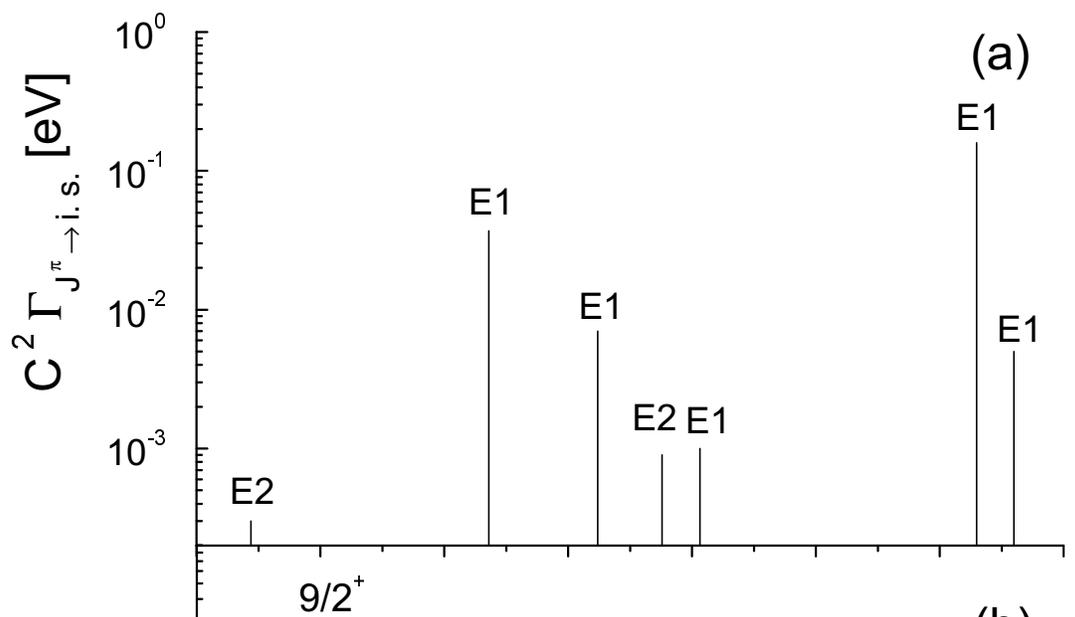
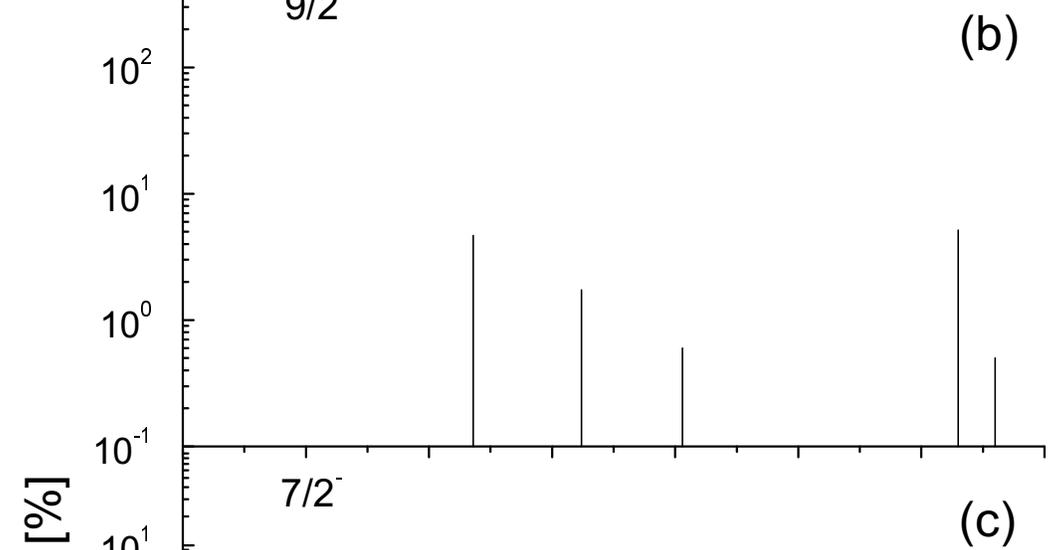
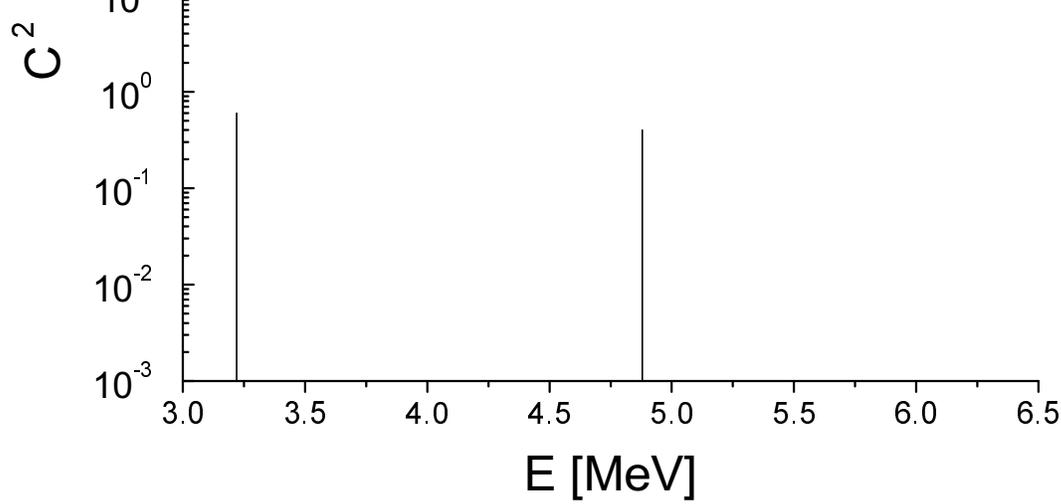
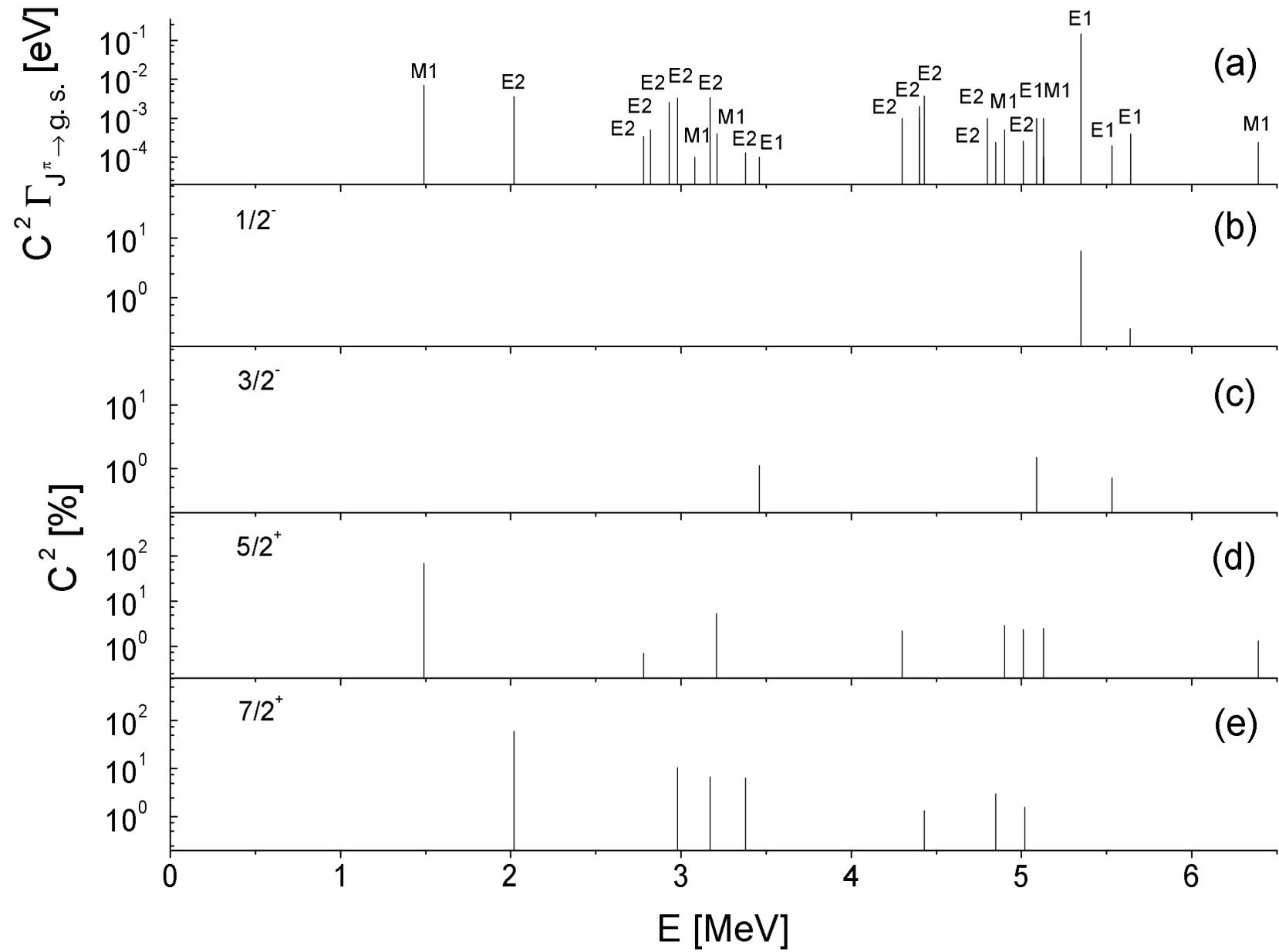
\end{document}